\documentclass{aa}
\usepackage{graphicx}
\usepackage{natbib}
\usepackage{xspace}
\usepackage[varg]{txfonts}
\usepackage{color}
\usepackage{url}
\usepackage{hyperref}

\newcommand{\lum}{erg s$^{-1}$}
\newcommand{\Nu}{{\it NuSTAR\xspace}}
\newcommand{\source}{GX 301$-$2\xspace}

\begin{document}

\title{\textit{NuSTAR} observations of the wind-fed X-ray pulsar GX 301--2 during an unusual spin-up event}

\author{ Armin Nabizadeh \inst{1} 
        \and Juhani M\"onkk\"onen \inst{1}
   		  \and Sergey S. Tsygankov \inst{1,2}
   		  \and Victor Doroshenko \inst{3,2}
   		  \and Sergey V. Molkov \inst{2}
          \and Juri Poutanen \inst{1,2,4}
          }
          
   \institute{Department of Physics and Astronomy, FI-20014 University of Turku, Turku, Finland \\ \email{armin.nabizadeh@utu.fi}
       \and
       Space Research Institute of the Russian Academy of Sciences, Profsoyuznaya Str. 84/32, Moscow 117997, Russia    
       \and
       Institut f\"ur Astronomie und Astrophysik, Universit\"at T\"ubingen, Sand 1, D-72076 T\"ubingen, Germany
       \and 
       Nordita, KTH Royal Institute of Technology and Stockholm University, Roslagstullsbacken 23, SE-10691 Stockholm, Sweden
          }
          
   \titlerunning{Transient disk in GX 301--2}
   \authorrunning{ Nabizadeh et al. }
   \date{June 2019}
   
\abstract{We report on \textit{NuSTAR} observations of the well-known wind-accreting X-ray pulsar \source\ during a strong spin-up episode that took place in January-March 2019. A high luminosity of the source in a most recent observation allowed us to detect a positive correlation of the cyclotron line energy with luminosity. Beyond that, only minor differences in spectral and temporal properties of the source during the spin-up, presumably associated with the formation of a transient accretion disk, and the normal wind-fed state could be detected. We finally discuss conditions for the formation of the disk and possible reasons for lack of any appreciable variations in most of the observed source properties induced by the change of the accretion mechanism, and conclude that the bulk of the observed X-ray emission is still likely powered by direct accretion from the wind.
}

\keywords{accretion, accretion disks -- magnetic fields -- pulsars: individual: GX 301--2 -- stars: neutron -- X-rays: binaries}

\maketitle

\section{Introduction} \label{sec:intro}

X-ray pulsars (XRPs) are binary systems in which a neutron star (NS) with strong magnetic field accretes matter from its companion star. Accretion flow is funneled by the magnetic field to the surface of spinning NS where pulsed X-ray emission is produced. Based on the dominant mass-transfer mechanism, two classes of XRPs are usually defined: wind-fed and disk-fed XRPs. The wind-accreting XRPs are fed directly by accretion of dense wind from their giant companions, whereas the disk-accreting systems transfer mass onto a NS through the Roche-lobe overflow or periodically from decretion disks of Be donors.  Under certain conditions, a transient accretion disk can also be formed in wind-fed systems as discussed, for example, recently by \citet{Karino2019}. The accretion mechanism defines the material torques affecting the NS, and the total torque either decelerating or accelerating the NS rotation. Given that accretion disks can only form when the accretion flow carries substantial angular momentum, disk accretion implies stronger accelerating torque, and so the appearance of an accretion disk can be manifested as a spin-up event \citep[see, for example,][]{Soker2004}.

\source\ (also known as 4U 1223$-$62) is a high mass X-ray binary system (HMXB) containing an XRP with period of $\sim$680\,s, one of the longest observed \citep{White1976}. In this system, accretion is normally fed by the wind from the hypergiant donor star, Wray 977. Spectral classification of the donor was used by \citet{Parkes1980} to measure the distance to the system of 1.8 kpc, however, later \citet{Kaper1995} obtained 5.3 kpc. On the other hand, \textit{Gaia} parallax measurement implies $d=3.53_{-0.52}^{+0.40}$~kpc \citep{Treuz2018,BailerJones2018}. This discrepancy might affect some of the derived properties of the companion, which is, however,
in any case a massive early B-type star with the mass of 39--53 $M_{\odot}$ and a radius of $\sim$62 $R_{\odot}$ \citep{Kaper2006}. The donor also exhibits extremely high mass loss rate $\dot{M} \sim 10^{-5} M_{\odot}$ yr$^{-1}$, one of the highest rates known in the galaxy. The wind is slow \citep[$\sim$300--400 km s$^{-1}$;][]{Kaper2006}, which allows the pulsar to capture it effectively thus reaching the X-ray luminosity of up to 10$^{37}$\lum. The orbital period of the system is $\sim$41.482$\pm$0.001\,d \citep{Doroshenko2010} with the orbit characterized by a high eccentricity of $e\sim$ 0.46 \citep{Koh1997}.



\begin{table*}
    \centering
    \caption{\textit{NuSTAR} observations of \source.}
    \begin{tabular}{ccccc}
    \hline
    ObsID & Start date & Start MJD & Exposure (ks) & Orbital phase  \\
    \hline
    30001041002 & 2014-10-29 & 56959 & 38.2 & 0.65 \\
    30101042002 & 2015-10-04 & 57299 & 35.7 & 0.85 \\
    90501306002 & 2019-03-03 & 58545 & 36.1 & 0.89 \\
    \hline
    \end{tabular}
    \label{tab:observations}
\end{table*}

\source\ exhibits regular periodic X-ray outbursts associated with orbital motion, $\sim$1.4 d before the periastron passage where the accretion rate is highest \citep{Pravdo2001}. A second peak in the orbital light curve occurs, however, near the apastron passage at the orbital phase $\sim$0.5 \citep{Pravdo1995,Koh1997} which is harder to explain. So far, several models have been proposed to interpret the observed
orbital light curve of the source. An inclined circumstellar disk feeding the accretion around
the periastron and apastron of Wray 977 was suggested \citep{Pravdo1995,Koh1997,Pravdo2001}. However, optical observations did not confirm  presence of such a disk \citep{Kaper2006}. Another possibility was proposed by \citet{Leahy2002} and   \citet{Leahy2008}, who argued that in addition to homogeneous stellar wind, a high density accretion stream trailing the NS along the orbit is present in the system. The stream origin is associated with tidal interactions, and the observed peaks in the orbital light curve are attributed to crossings of the stream by the pulsar near the periastron and apastron. Spectral analysis of the optical companion found the evidence for such a spiral-type stream in this system \citep{Kaper2006}, moreover, later mid-infrared interferometry confirmed the presence of the stream in the system \citep{Inferared-Gravity2017}.

The X-ray spectrum of \source\ has been reported to exhibit two absorption features at 34 and 51~keV  \citep{Furst2018} interpreted as Cyclotron Resonant Scattering Features (CRSFs), although earlier investigations found a single pulse-phase-dependent feature around 30--50 keV \citep{Kreykenbohm2004,LaBarbera2005}, so there is some
ambiguity in interpretation of the spectrum. 
The variation of the CRSF energy with pulse and orbital phase had been a topic of several investigations
\citep{Kreykenbohm2004,LaBarbera2005,Doroshenko2010}. The reason is that in several XRPs, the centroid energy of the CRSF has been reported to change with
luminosity \citep[see e.g.][]{Mihara1998,2019A&A...622A..61S}. In particular, in low luminosity XRPs, the energy of the line seems to be correlated with
flux \citep{Staubert2007,Klochkov2012}, whereas at higher accretion rates the correlation tends to be negative \citep{Tsygankov2006,2006ApJ...646.1125N,Tsygankov2010}. The change in
luminosity-dependency has been connected to the source exceeding the critical luminosity when the accretion column starts to rise
\citep{Basko1976,Becker2012,2015MNRAS.447.1847M}. Recently this scenario has been
confirmed observationally when both types of behaviour with the transition at expected luminosity were observed in V~0332+53 \citep{Doroshenko2016}.

\begin{figure}
\begin{center}  
\includegraphics[width=0.9\columnwidth]{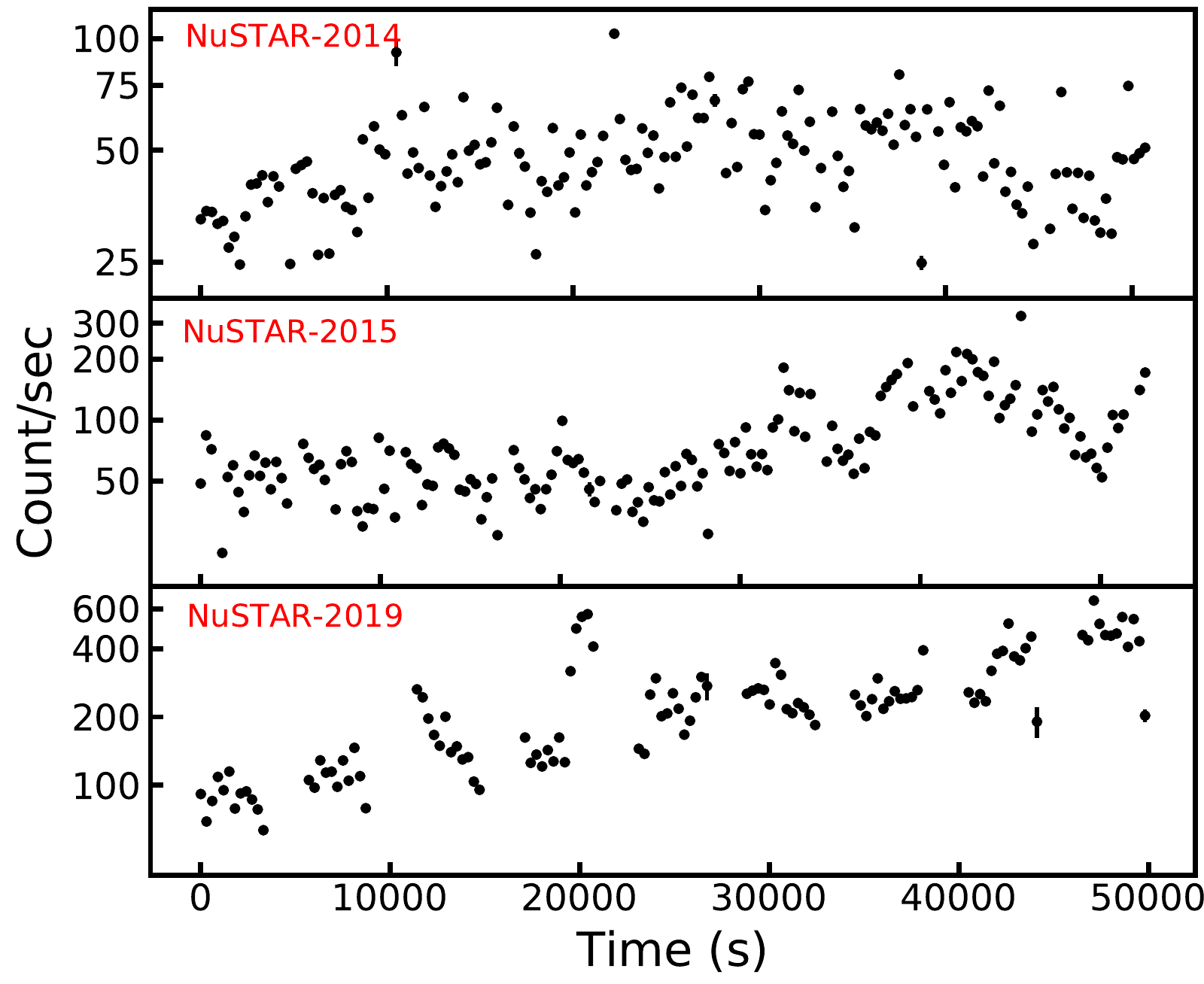}
\end{center}
\caption{X-ray light curves of \source\ extracted from the three \textit{NuSTAR} observations in the energy range  3--79 keV.} 
\label{fig:ltc} 
\end{figure}

\citet{LaBarbera2005} studied the X-ray spectrum of \source\ over an orbital cycle of the binary and found no evidence of variations of the CRSF with luminosity, pointing to sub-critical accretion. Later, \citet{Suchy2012} reported a hint of anti-correlation, although also in this case no definite conclusion on the accretion regime could be made. Here we report on recent \Nu\ observation of the source during a rapid spin-up episode when luminosity of the source was higher than usual (see Fig.~\ref{fig:ltc}). Several rapid spin-up episodes associated with formation of a transient accretion disk have been reported for \source\ \citep{Koh1997,Bildsten1997}, however, until now, no dedicated broadband observations have been carried out in this state. Using the archival \Nu\ observations of the source in normal state as a baseline, we investigated, therefore, how the presence of an accretion disk affects the temporal and spectral properties of the source. We re-visit the issue of luminosity dependence of the CRSF and discuss how these irregular spin-up events can be related to the apastron passages.

\section{Observations and data reduction} \label{sec:data}

\begin{figure*}
\begin{center}
\includegraphics[width=0.95\textwidth]{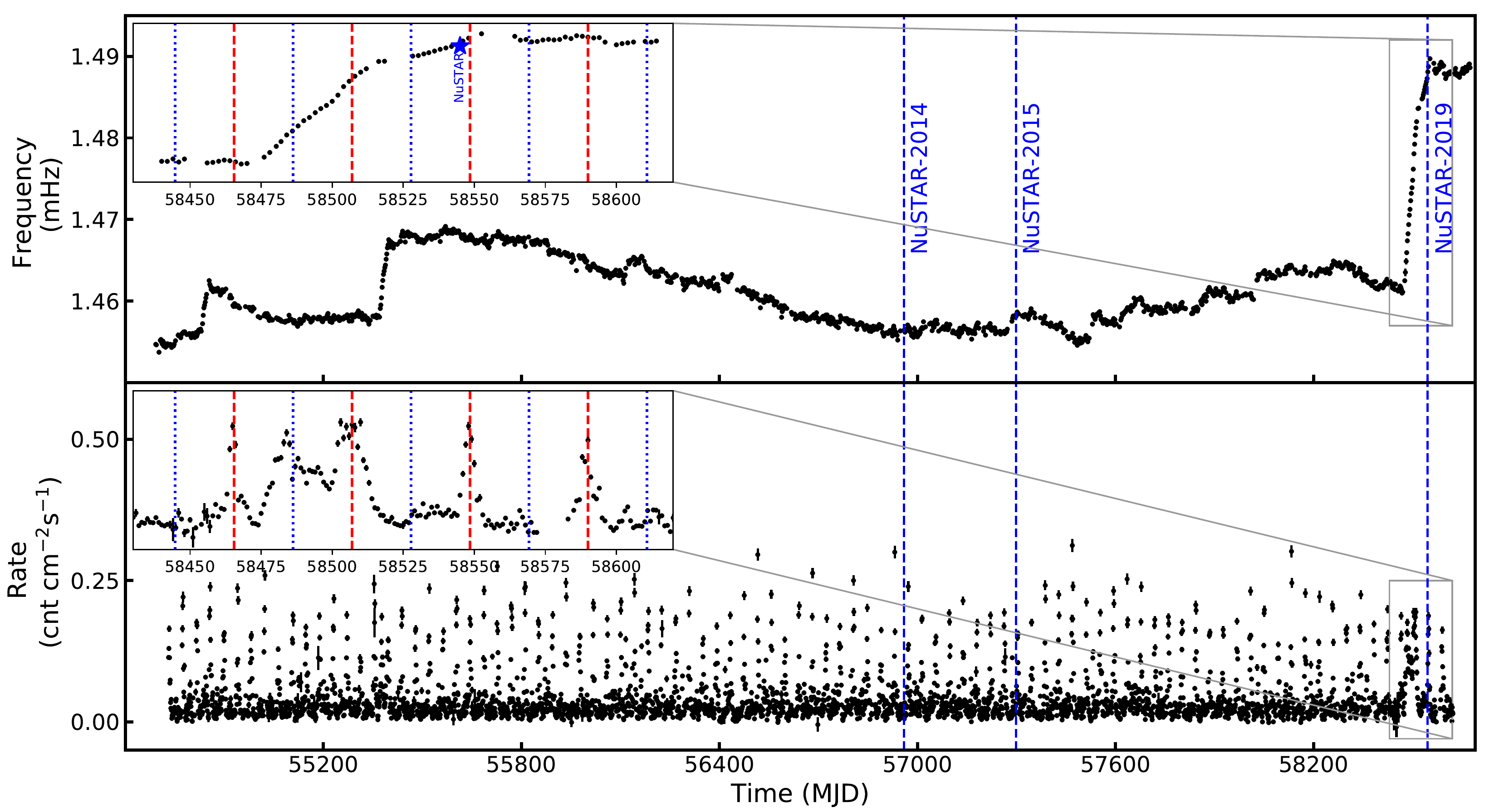}
\end{center}
\caption{Upper panel: The pulse frequency of \source\ observed with the \textit{Fermi} Gamma-ray Burst Monitor detectors (GBM Pulsar Project). Lower panel: X-ray count rate seen by \textit{Swift}/BAT in the energy range 15--50 keV. Blue dashed lines show the dates when the \Nu\ observations were performed. In each panel the recent spin-up episode is zoomed-in, with the periastron passages marked with the red dashed lines and the apastron passages shown with the blue dotted lines.
} 
\label{fig:bat} 
\end{figure*}

The \Nu\ observatory consists of two co-aligned grazing incidence X-ray telescope systems (FPMA and FPMB) with independent CdZnTe detector units operating in a wide energy range of 3--79~keV \citep{Harrison2013}. It provides X-ray imaging resolution of 18\arcsec\ (full width at half-maximum, FWHM) and spectral resolution of 400 eV (FWHM) at 10\,keV. 
\source\ has been observed by \Nu\ three times on October 2014, October 2015 and March 2019 (see Table~\ref{tab:observations} for more details). The first observation (ObsID 30001041002) was performed soon after the apastron passage while the second one (ObsID 30001042002) was scheduled at intermediate phase between apastron and periastron. The most recent ToO observation is the main topic of current investigation (ObsID 90501306002; PI M\"onkk\"onen), and was performed on 2019 March 3 with an exposure time of 36.1~ks. This observation specifically targeted the rare rapid spin-up episode (MJD 58475--58550), and is the first dedicated observation of \source\ in such state. The exact orbital phases corresponding to \Nu\ observations are listed in Table~\ref{tab:observations} and indicated in Fig.~\ref{fig:ratio}. Using the standard data reduction procedure explained in the \Nu\ user guide\footnote{\url{https://nustar.ssdc.asi.it/news.php\#}}, we reduced the data and extracted the source spectra and the light curves using the \Nu\ Data Analysis Software {\sc nustardas} v1.8.0 with a {\sc caldb} version 20180419. To extract the data products we used source-centered circular region with radius of 120\arcsec\ for both FMPA and FMPB. The background was extracted similarly from a source-free circular region of the same radius in the corner of the field of view. We then applied the barycentric correction to all the resulting light curves using the standard {\sc barycorr} task. In order to increase the signal to noise ratio and ensure  good statistics in the spectral bins, the spectra were grouped to have at least 70 counts in each energy bin. The spectra of the source measured by the two \Nu\ units were fitted independently as recommended in instruments documentation.

\section{Analysis and results} \label{sec:timing}

\begin{figure}
\begin{center}
\includegraphics[width=0.95\columnwidth]{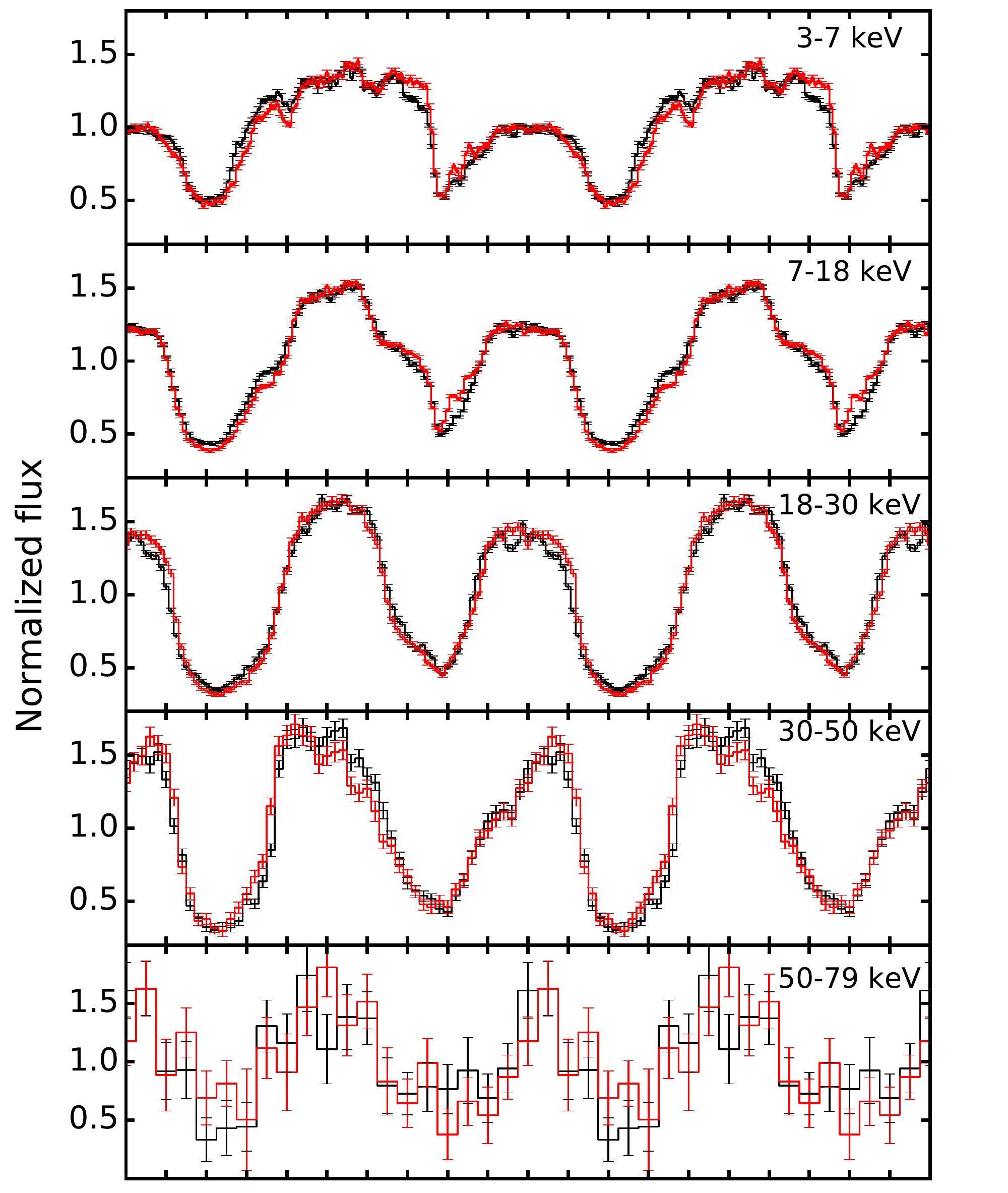}
\includegraphics[width=0.95\columnwidth]{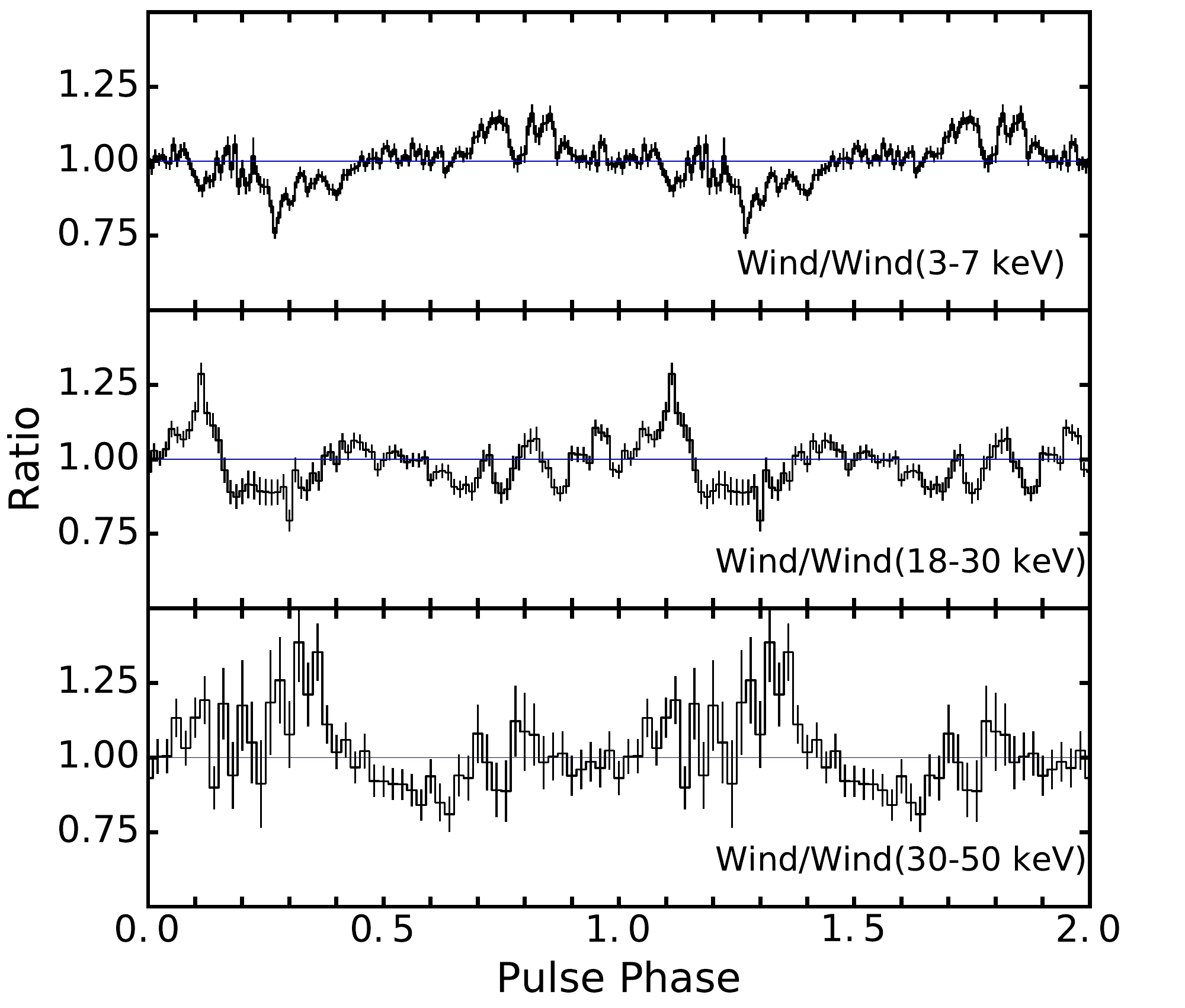}
\end{center}
\caption{\textit{Top panels}: The pulse profiles of \source\ in five different energy bands 3--7, 7--18, 18--30, 30--50 and 50--79 keV (from top to bottom) obtained by \Nu\ observations in 2014 (red) and 2015 (black) during the wind accretion phases. The fluxes are normalized by the mean flux in each energy band.  The zero phase was chosen to maximize the  CCF.  
\textit{Bottom panels}: Ratio of the \source\ pulse profiles during the wind-fed states (wind/wind) obtained using 2014 and 2015 observations in the energy bands 3--7, 18--30 and 30--50 keV. The ratio of unity is shown by the  horizontal blue line.} 
\label{fig:pp1} 
\end{figure}

\begin{figure}
\begin{center} 
\includegraphics[width=0.95\columnwidth]{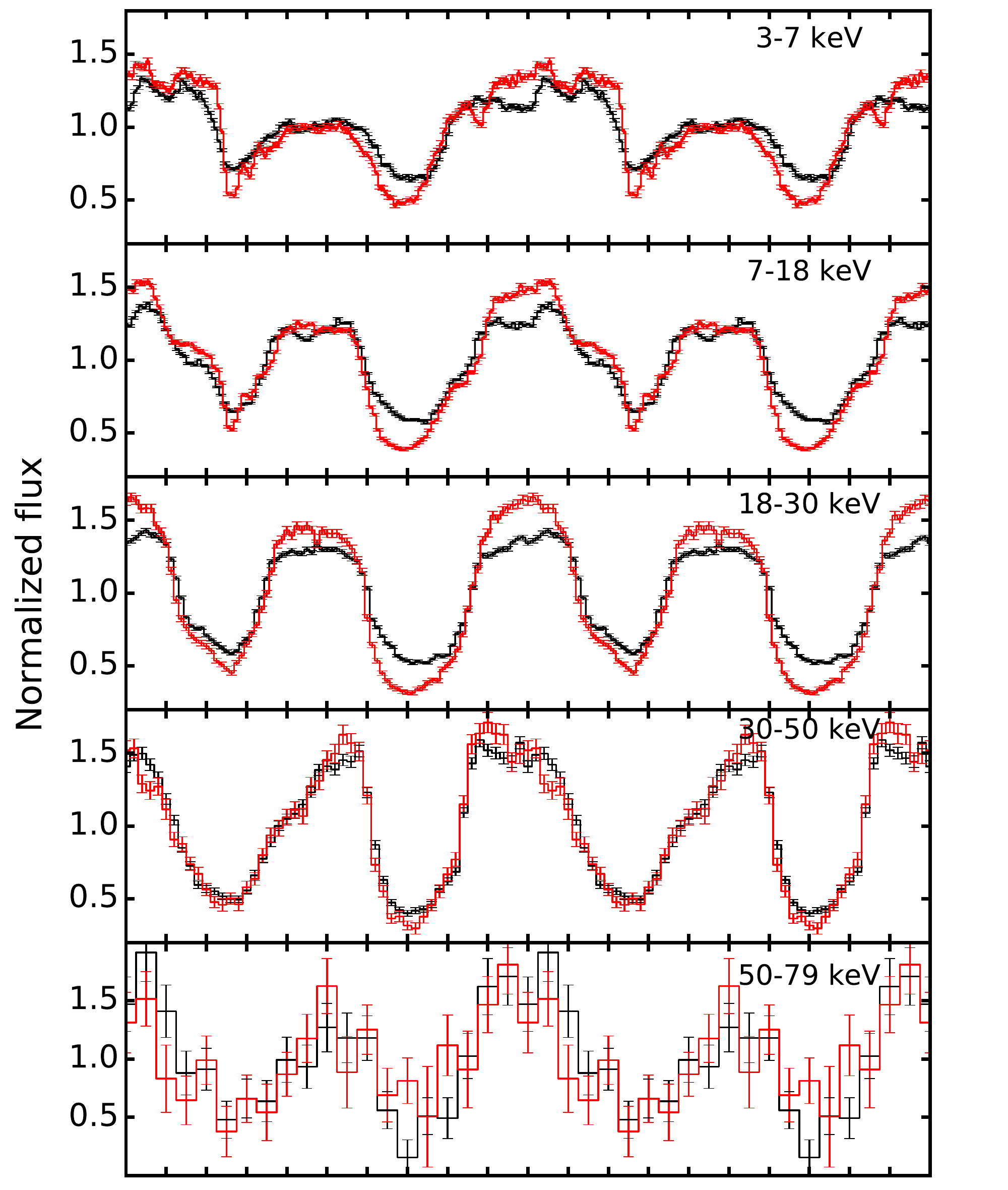}
\includegraphics[width=0.95\columnwidth]{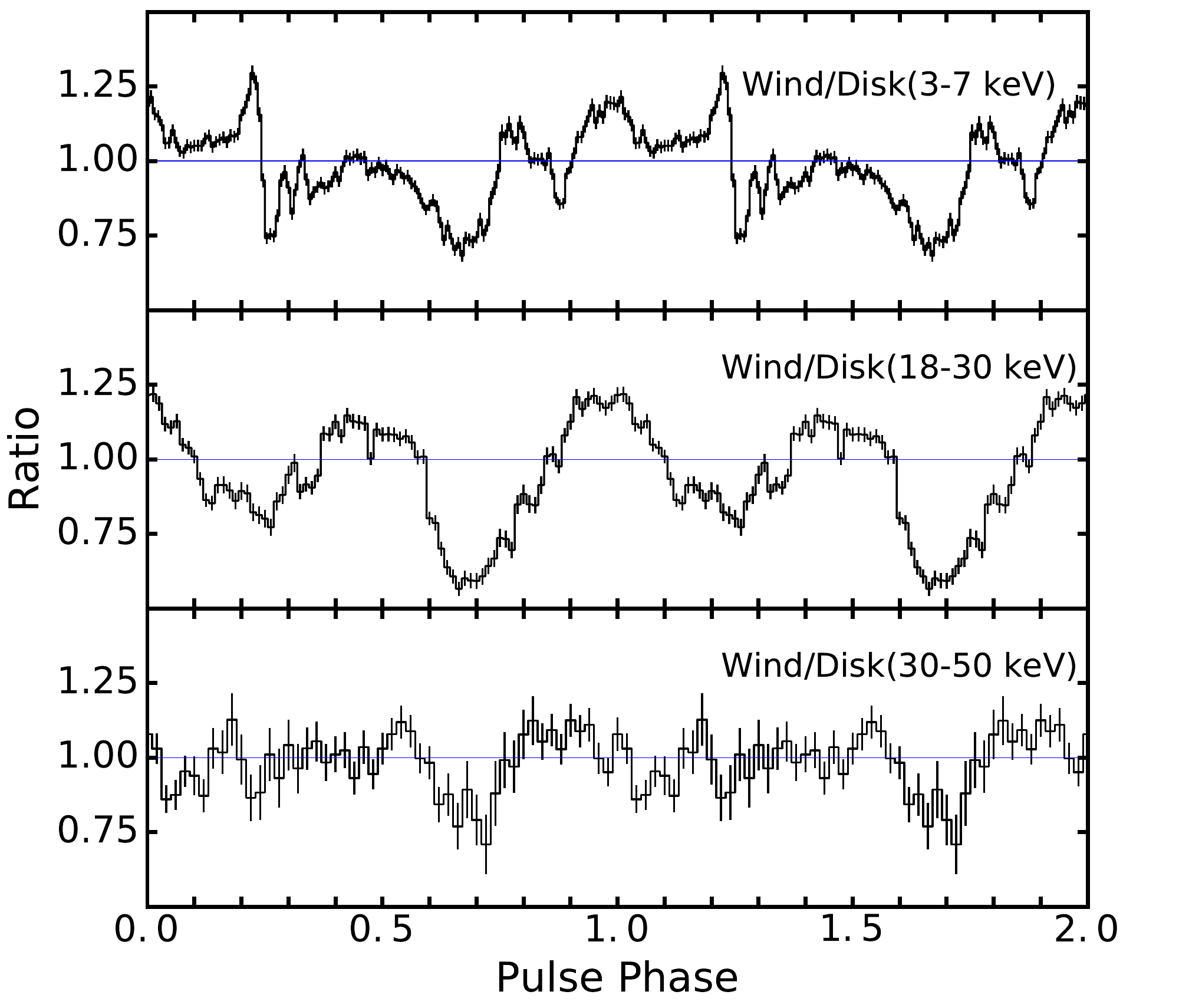}
\end{center}
\caption{Same as Fig.\,\ref{fig:pp1}, but for the observations in 2014 (red) and 2019 (black) in which the NS was accreting from the wind and the disk, respectively.}
 
\label{fig:pp2} 
\end{figure}
\subsection{\Nu\ observation of the spin-up episode. }

In March 2019, the \textit{Fermi} Gamma-ray Burst Monitor detectors (GBM Pulsar Project\footnote{\url{http://gammaray.nsstc.nasa.gov/gbm/science/pulsars/}}) detected a rapid spin-up episode for \source, which is the fourth such episode in history of observations of the source, and the most powerful in recent years. The change of the pulse frequency as observed by \textit{Fermi}/GBM \citep{2009ApJ...702..791M} and the simultaneous count rate obtained by \textit{Swift}/BAT \citep{2013ApJS..209...14K} in the energy range 15--50  keV\footnote{\url{https://swift.gsfc.nasa.gov/results/transients/GX301-2/}} are shown in Fig.~\ref{fig:bat}. 

Prior to the spin-up episode the source exhibited an initial pre-periastron flare on MJD 58465, followed by an approximately 10-day flux decay. On MJD 58475, the flux started to rise again at phase $\sim0.25$ marking the start of the spin-up episode. This secondary flare peaked at phase $\sim0.5$ on MJD 58484, the first day of 2019, when the pulsed flux also rose suddenly. The flux remained elevated until a multi-peaked pre-periastron flare took place from MJD 58503 to 58510 (around the expected flare on MJD 58507). The average spin-frequency derivative during the strong spin-up was $0.2$~mHz~yr$^{-1}$. The flux dropped after the periastron passage, which affected the spin frequency evolution. In a period from MJD 58527 to 58548 between apastron and periastron passages, the flux was slightly above average, and the source exhibited spin-up with a lower average derivative of $0.07$~mHz~yr$^{-1}$. Our \Nu\ observation took place during this period on MJD 58545, right before the source reached the pre-periastron flare around MJD 58548, i.e. close to the end of the most recent spin-up episode lasting $\sim70$\,d with pulse period decreasing in total from 684~s to 672~s.

As seen in Fig.\ref{fig:ltc}, the object was significantly brighter and varied on a larger scale during the latest \Nu\ observation compared to the previous two observations with the same instrument. 
In particular, the mean count rate increased by $\sim$200 cnt s$^{-1}$, and a strong flare/outburst $\sim$20 ks after the beginning of observation lasting for a couple of pulse periods was observed. Given that the three \Nu\ observations took place at different orbital phases, besides the fact that the source is more variable near the periastron, the higher flux level is probably not directly related to the spin-up episode.

\subsection{Pulse profile and pulsed fraction}
\label{sec:pulse}
For timing analysis we corrected light curves for the binary motion using the orbital parameters reported by \citet{Koh1997} and \citet{Doroshenko2010} as $P_{\rm orb}$ = 41.472\,d, $a_{\rm x}$sin$i$ = 368.3 lt s, $\omega$ = 310\fdg4, $T_{\rm periastron}$ = MJD 53531.65 and $e$=0.462. To estimate the pulse period for each observation, we used the standard epoch folding technique \citep{1983ApJ...266..160L} using the {\sc efsearch} procedure from the {\sc ftools} package which yielded the spin periods as $P_{2014}=686.47(3)$~s, $P_{2015}=685.94(1)$~s and $P_{2019}= 672.51(5)$~s. Uncertainties were determined from simulations of large number of light curves using count rates and corresponding error bars from the original data \citep[for details, see ][and references therein]{2004AstL...30..824F}.

The wide energy coverage of \Nu\ allows to study the energy dependence of the pulse profiles. Considering the available counting statistics, we extracted the
light curves in five different energy ranges 3--7, 7--18, 18--30, 30--50 and 50--79 keV from all three observations.
We verified that light-curves extracted from two modules are consistent with each other, and therefore, co-added the data from both detectors to improve counting statistics. Using the task {\sc efold} from {\scriptsize XRONOS} package, we folded
the energy-dependent light curves of each observation with the corresponding pulse period. 

First, we compare the pulse profiles during the wind-fed accretion states in Fig.~\ref{fig:pp1}. The zero phases are chosen to maximize the cross-correlation function (CCF). 
Then, we also compare the pulse profiles obtained in 2019 during the disk-fed state  to that obtained in 2014 in the wind-fed state in Fig.~\ref{fig:pp2}. 
Five top panels in  these two figures show the evolution of the pulse profiles with energy increasing from top to bottom. In all three
observations the pulse profiles are double-peaked with complex substructures presented in lower energy bands and tendency for the shape simplification at higher energies. In particular, the broad main peak in 3--7 keV is getting narrower at higher energies. In order to
better emphasize the difference between the pulse profiles observed in wind-fed and disk-fed states, we show their ratios for 2014 and 2015 observations (wind/wind) and for 2014 and 2019 observations (wind/disk) in three energy bands 3--7, 18--30 and 30--50 keV in the
three bottom panels of Figs.~\ref{fig:pp1} and ~\ref{fig:pp2}. The low amplitude variations of the ratio in wind/wind scenario confirms that \source
has a similar geometrical configuration when accreting from the wind.  Although stronger phase dependence of the wind/disk ratio is apparent from the bottom panels of Fig.~\ref{fig:pp2}, the overall structure and energy dependence of the pulse profiles in different accretion regimes suggest generally similar structure of the emission region in both cases.

\begin{figure}
\begin{center}    
\includegraphics[width=0.9\columnwidth]{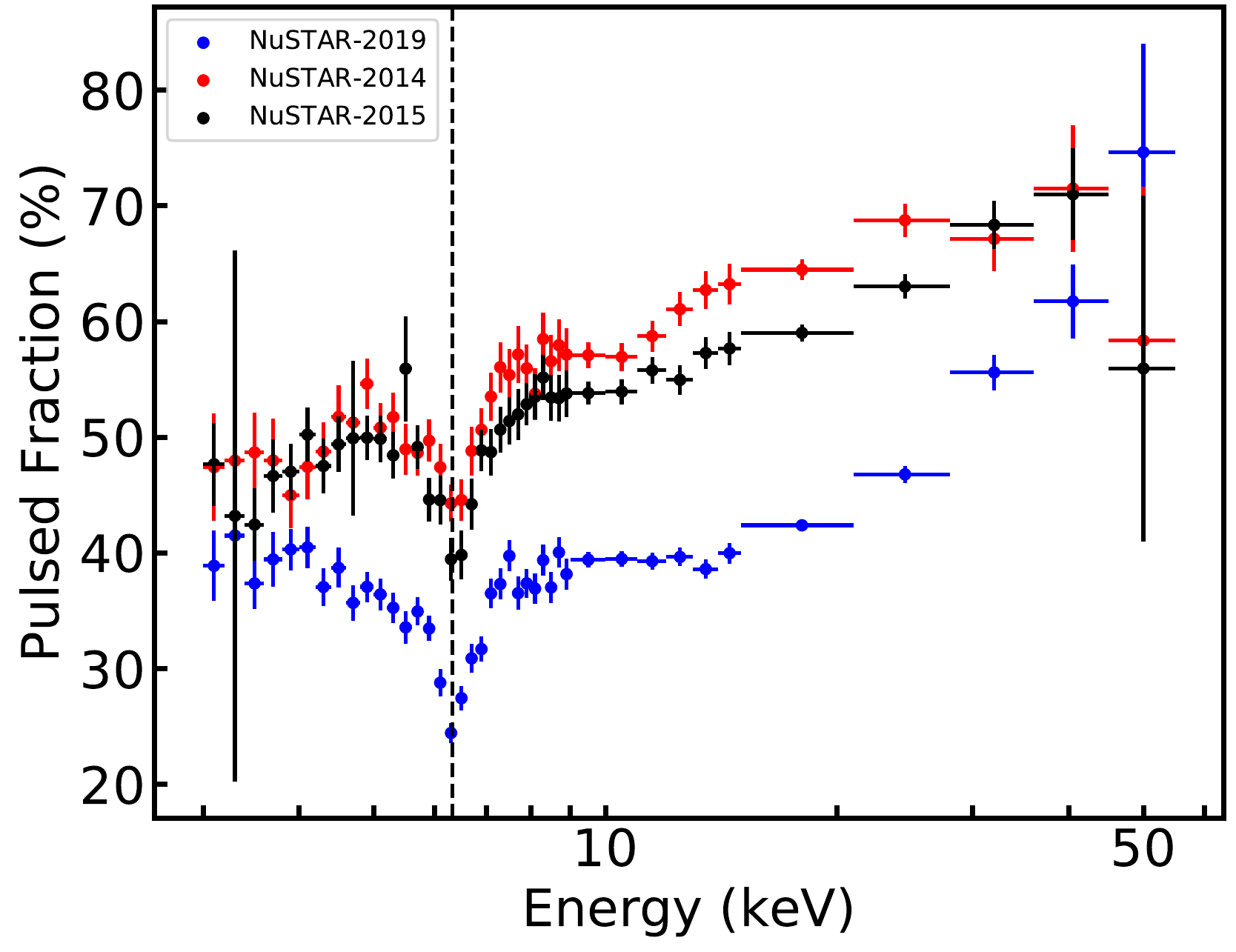}
\end{center}    
\caption{The energy dependence of the pulsed fraction of \source\ obtained from the \textit{NuSTAR} observations in 2014 (red crosses), 2015 (black crosses) and 2019 (blue crosses). The vertical dashed line indicates the centroid of the iron K$\alpha$ emission line. } 
    \label{fig:pf}
\end{figure}

This difference is also captured by energy dependence of the pulsed fraction defined as PF = $(F_{\rm max} - F_{\rm min})/(F_{\rm max} + F_{\rm min})$ (here $F_{\rm min}$ and $F_{\rm max}$ are the minimum and maximum fluxes of the pulse profile) as illustrated in Fig.~\ref{fig:pf}. While in all three observations the pulsed fraction generally increases with energy as observed for the majority of XRPs \citep{Lutovinov2009}, the pulsed fraction in the
hard band (8--20 keV) appears to be significantly lower during the spin-up phase. In addition, we observe a sharp decrease of the pulsed fraction associated with the iron K$\alpha$ emission line in the 6--7 keV range.

\section{Spectral analysis} \label{sec:spectral}

The spectrum of \source, as typical for other accreting XRPs, is believed to be due to Comptonization process in the vicinity of a NS. Such a continuum has
power-law shape with a cut-off at the energy related to the temperature of the Comptonizing region. Therefore, in the early studies of the source, a modified high
energy cut-off \citep{LaBarbera2005} and Fermi-Dirac cut-off \citep{Kreykenbohm2004} were used.
Regardless on the assumed intrinsic spectrum, inclusion of a partial-covering absorber was found to be required to account  for both interstellar absorption and absorption within the binary system \citep{Kreykenbohm2004,LaBarbera2005}.  

\citet{Kreykenbohm2004}, using the \textit{RXTE} data, confirmed that there is a broad CRSF at 35~keV which was firstly detected by \citet{Mihara1995}. The
authors found that the energy of the line is strongly variable with pulse phase changing between 30 and 40~keV. Later, \citet{Suchy2012} associated the observed
sinusoidal variations of the line energy with the pulse phase with the change of the angle between the line of sight and the magnetic field of the NS.
\cite{LaBarbera2005} studied \textit{BeppoSAX} observations and detected a CRSF at two different energies, which changed with orbital phase, i.e. 45 and 53
keV for periastron and pre-periastron, respectively. 
Later, using Comptonization model {\sc compst} by \citet{SunyaevTitarchuk1980} to describe the continuum, \citet{Doroshenko2010}  detected a CRSF at $\sim$46 keV in the \textit{INTEGRAL} data, confirming the values reported by \citet{LaBarbera2005}.

In 2018, \citet{Furst2018} published a detailed spectral analysis of \source\ using the two \Nu\ observations in 2014 and 2015 when the source was accreting from the wind (the same observations we used here). They applied several phenomenological models such as {\sc npex} \citep{Mihara1998}, {\sc fdcut} \citep{Tanaka1986} and  {\sc highecut} together with a partial covering absorber model to fit the spectra. Two gaussian absorption features associated with the CRSF were also required to model the spectrum. The energies of the two features at $\sim$35 and $\sim$50 keV appear to be not harmonically related because no features around 17 keV were found. \citet{Furst2018} suggested that both features are likely associated with fundamental cyclotron line, but originate at two different altitudes, e.g. at the NS surface and 1--1.4 km above the surface in the accretion column. Here we use similar approach to facilitate the comparison between the disk and the wind-fed states.

\begin{table*}
\begin{center}
	\caption{Best-fit model parameters for the \textit{NuSTAR} averaged spectra of 2019, 2015 and 2014.}
	
	\label{tab:2}
	\begin{tabular}{lccccr} 
		\hline
		Model & Parameters & Units                                 & 2019                       & 2015                       & 2014\\
		\hline
	\textsc{constant}$^{a}$ 	&  &                           & 1.017$\pm$0.001            & 1.035$\pm$0.001            & 1.035$\pm$0.002\\
        		 
        \textsc{tbabs}    &  $N_{\rm H,1}$         & $10^{22}$ cm$^{-2}$   &  1.4(Fixed)                & 1.4(Fixed)                 &  1.4(Fixed) \\
		\textsc{tbpcf}    &  $N_{\rm H,2}$         & $10^{22}$ cm$^{-2}$   &  51.1$^{+0.4}_{-0.7}$      & 31.6$^{+0.8}_{-0.8}$       &  39.7$\pm$0.7 \\
				 &  pcf                 & 	                  &  0.842$^{+0.002}_{-0.002}$ & 0.747$^{+0.005}_{-0.005}$     &  0.879$\pm$0.003 \\

	 \textsc{gabs}     &  $E_{\rm CRSF1}$          & keV	                &  55.1$^{+0.7}_{-0.6}$        & 50.1$^{+0.8}_{-0.8}$       & 49.6$^{+1.2}_{-1.3}$ \\
				 &  $\sigma_{\rm CRSF1}$     & keV	                &  4.2$^{+0.5}_{-0.4}$         & 6.5$^{+1.5}_{-1.0}$        & 9.8$^{+4.7}_{-1.9}$  \\
                 &  $D_{\rm CRSF1}$          &	                    &  8.5$^{+1.6}_{-1.4}$         & 17.1$^{+3.3}_{-4.1}$       & 37.0$^{+19.9}_{-5.9}$  \\
	\textsc{gabs}    &  $E_{\rm CRSF2}$          & keV	                &  36.7$^{+0.3}_{-0.3}$        & 35.8$^{+1.0}_{-1.2}$       & 34.6$^{+2.8}_{-0.6}$   \\
				 &  $\sigma_{\rm CRSF2}$     & keV	                &  11.8$^{+0.3}_{-0.4}$        & 6.4$^{+0.5}_{-0.5}$        & 6.2$^{+0.5}_{-0.7}$   \\
                 &  $D_{\rm CRSF2}$          &	                    &  42.1$^{+2.5}_{-3.4}$        & 10.6$^{+2.4}_{-2.3}$        & 7.1$^{+7.6}_{-3.2}$   \\
	\textsc{bbodyrad}    &  $kT$             & keV	                &  1.409$^{+0.006}_{-0.005}$   & 1.25$^{+0.02}_{-0.02}$     & 1.39$^{+0.08}_{-0.08}$ \\
				         &  $A_{\rm BB}$     & $\times10^{-2}$       &  40.6$^{+0.8}_{-1.0}$      & 7.9$^{+0.6}_{-0.6}$     & 1.16$^{+0.01}_{-0.01}$ \\
	\textsc{fdcut} &  $\Gamma$            & 	                    &  0.25$^{+0.02}_{-0.06}$      & 1.07$^{+0.05}_{-0.15}$     & 1.38$^{+0.02}_{-0.10}$ \\
                    &  $E_{\rm cut}$       & keV	                &  30.5$^{+0.3}_{-0.6}$        & 31.0$^{+2.25}_{-0.8}$       &  44.5$^{+16.1}_{-12.4}$\\
		    		 &  $E_{\rm fold}$      & keV	                &  5.04$^{+0.08}_{-0.19}$      & 7.9$^{+0.4}_{-0.3}$        &  6.9$^{+0.8}_{-3.0}$\\
		    		 &  $A_{\rm PL}$            &                       &  0.04$^{+0.07}_{-0.09}$      & 0.08$^{+0.01}_{-0.02}$      & 0.118$^{+0.9}_{-0.3}$ \\
    \textsc{gaussian} &  $E_{\rm Fe}$        & keV	                &  6.3206$\pm$0.0001           & 6.3594$^{+0.0002}_{-0.0003}$          & 6.3596$^{+0.0003}_{-0.0004}$ \\
				 &  $\sigma_{\rm Fe}$   & keV	                &  0.001(fixed)                & 0.001(fixed)               & 0.001(fixed) \\
                 &  $A_{\rm Fe}$ & $10^{-3}$ ph s$^{-1}$cm$^{-2}$   &  5.62$^{+0.05}_{-0.05}$  & 1.4$^{+0.02}_{-0.02}$      & 0.7$\pm$0.02\\
          $F_{\rm 3-79}$&	 & $10^{-9}$ erg s$^{-1}$cm$^{-2}$   & 8.251$^{+0.007}_{-0.006}$   & 2.513$^{+0.003}_{-0.003}$  & $1.617^{+0.002}_{-0.003}$\\
		 $\chi^2_{\rm red}$(d.o.f.)  & 	&	                        & 1.2(1808)                    & 1.00(1619)                & 1.08(1568) \\ 
    \hline
		 
	\end{tabular}
\end{center}
     $^{a}$Cross-calibration normalization factor between FPMA and FMPB instruments on-board \Nu. 
\end{table*}

\subsection{Phase-averaged spectroscopy}

We used all three \Nu\ data sets to investigate the evolution of the spectrum using {\sc xspec} version 12.9.1p \citep{1996ASPC..101...17A}. In order to perform the phase-averaged spectral analysis, the spectra were extracted using the procedure described  in Sect.~\ref{sec:data}. Following \citet{Furst2018}, to describe the spectral shape we considered a power-law model with a Fermi-Dirac cutoff  \citep[{\sc fdcut};][]{Tanaka1986}:
\begin{equation}
    F(E) = A_{\rm PL} \frac{E^{-\Gamma}}{{\rm exp}((E - E_{\rm cut})/E_{\rm fold}) + 1} 
\end{equation}
where $A_{\rm PL}$ and ${\rm \Gamma}$ are the power-law normalization constant and the photon index, $E_{\rm cut}$ and $E_{\rm fold}$ are the cutoff and the folding energies of the Fermi-Dirac cutoff.
Similarly to earlier reports we found also that two-component absorption model
was required to describe the spectrum with two column densities $N_{\rm H,1}$, responsible for interstellar, and $N_{\rm H,2}$ for the local absorption, respectively. In practice, we modeled the absorption using {\sc tbabs}, the Tuebingen-Boulder interstellar medium absorption model \citep{2000ApJ...542..914W} together with a partial covering  absorption ({\sc tbpcf}) to determine the corresponding values of $N_{\rm H,1}$ and $N_{\rm H,2}$. Interstellar absorption can be estimated based on radio data using the online tool  {\sc nhtot}\footnote{\url{http://www.swift.ac.uk/analysis/nhtot/}} \citep{Willingale2013} where the mean value of the hydrogen column density in the direction of \source\ is reported as 1.4$\times10^{22}$ cm$^{-2}$. We fixed $N_{\rm H,1}$ at this value and let $N_{\rm H,2}$ to vary.

Similarly to \cite{Furst2018}, we also found residuals in the soft band which we accounted for by inclusion of a  blackbody component ({\sc bbodyrad} in {\sc xspec}) with the temperature of $\sim1.4$\,keV which significantly improved the fit. \citet{Furst2018}, on the other hand, associated the soft-band residuals with slight miscalibration of \Nu\ gain, and added a gain shift to the FMPA spectrum finding discrepancy of $\sim$20 eV in gain between the \Nu\ instruments. 
In any case, either approach does not affect the spectra in the hard band, which are of our interest.

\begin{figure}
\begin{center}  
\includegraphics[width=\columnwidth]{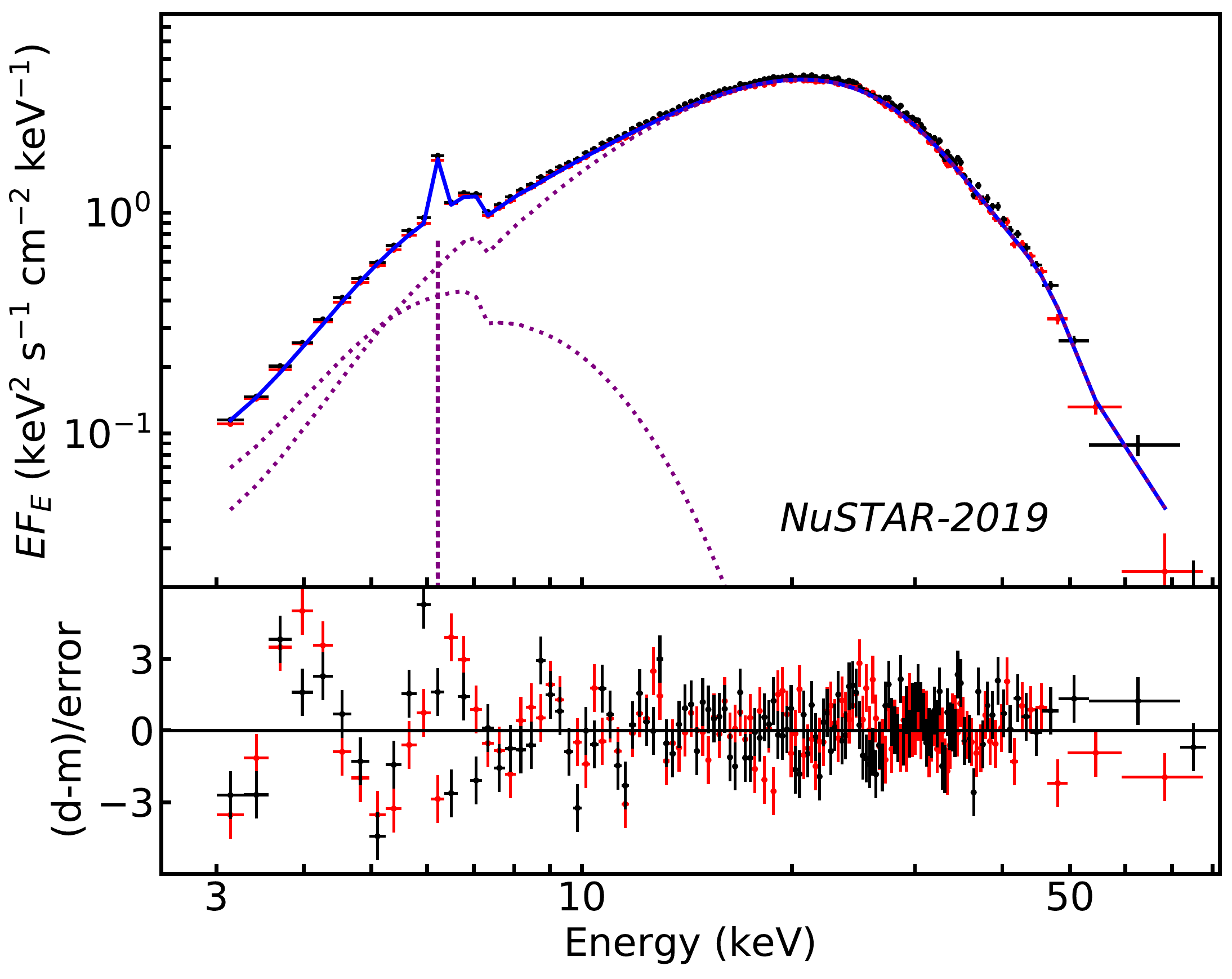}
\includegraphics[width=\columnwidth]{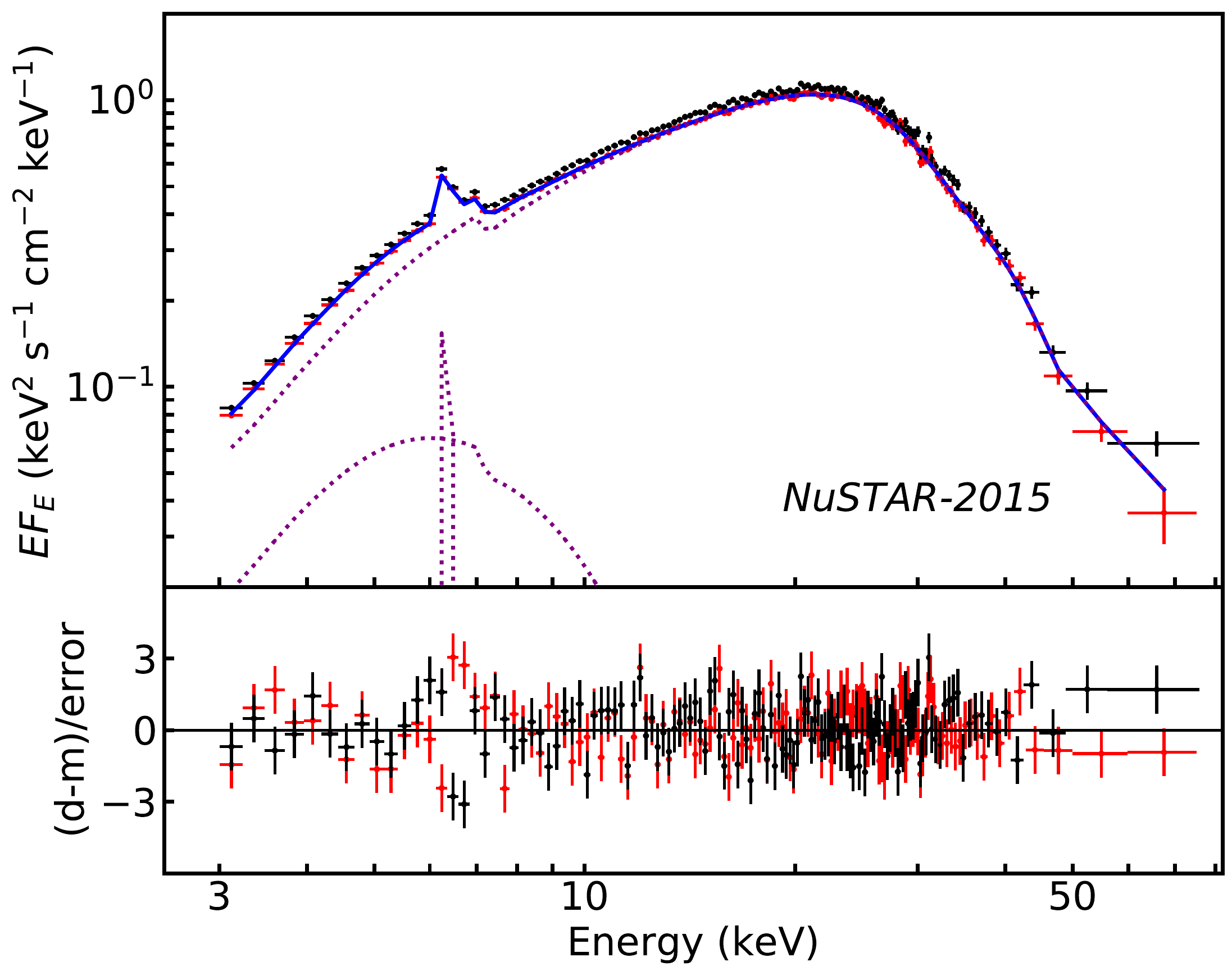}
\includegraphics[width=\columnwidth]{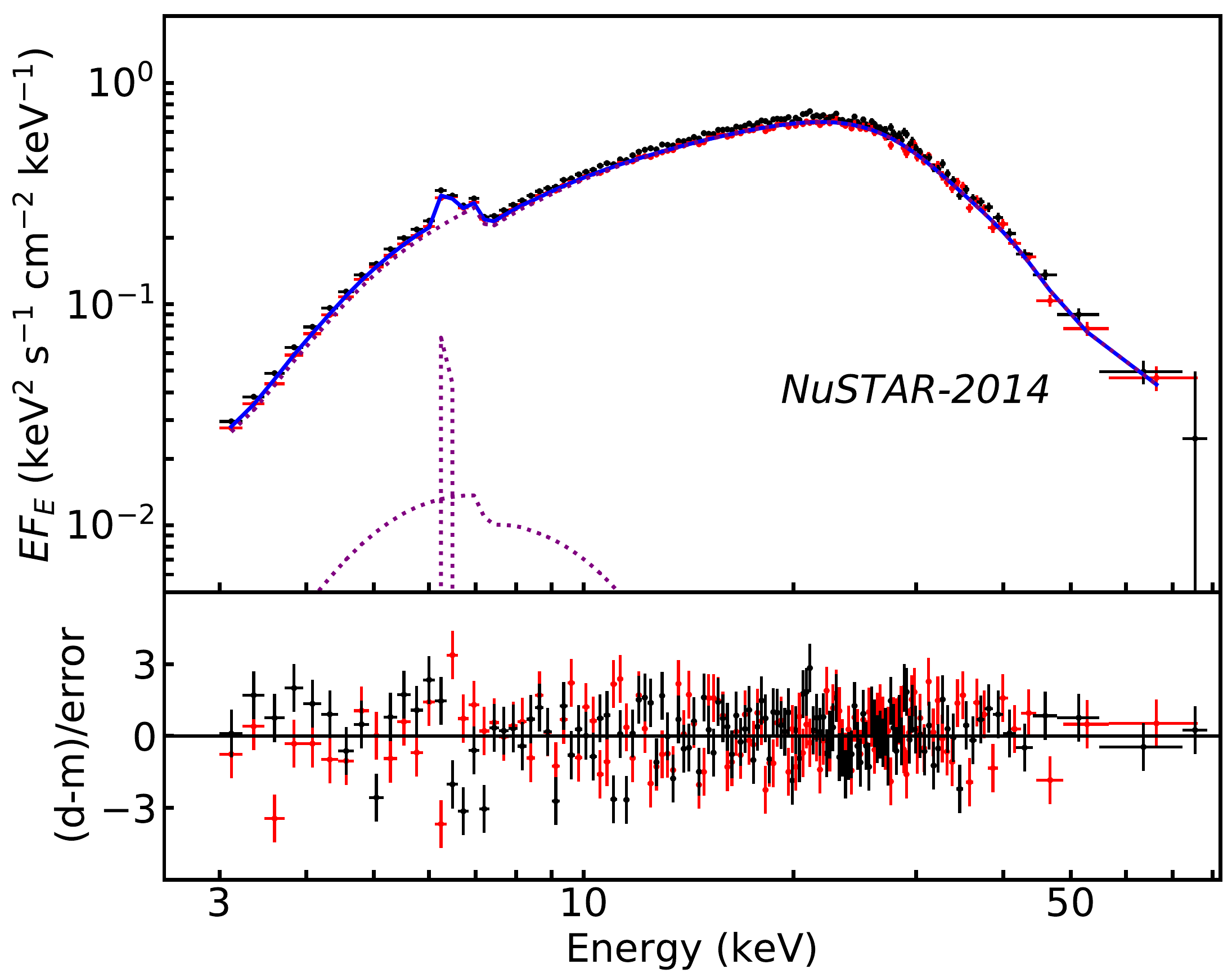}
\end{center}
\caption{Upper panels in each plot show the broad-band spectrum of \source\ obtained by FPMA and FPMB/\Nu\ (red and black crosses) in 2019, 2015 and 2014 together with the best-fit model {\sc constant $\times$ tbabs $\times$ tbpcf $\times$ gabs $\times$ gabs $\times$ (bbodyrad + fdcut + gaussian)} (solid line). Different continuum components and the iron line are shown with dashed lines.  Bottom panels show residuals from the best-fit model in units of standard deviations.} 
\label{fig:spec} 
\end{figure}

We added two Gaussian absorption line models ({\sc gabs} in {\sc xspec}) to account for the CRSF lines previously detected for the source. As expected the residuals were improved significantly. We also used a {\sc gaussian} component to account for the iron fluorescent emission line; it has a peak at 6.32 keV and normalization $A_{\rm Fe}$ = 5.6 $\times$ $10^{-3}$ ph s$^{-1}$ cm$^{-2}$. We keep the width frozen at 0.001 keV. These line parameters are consistent with other X-ray analyses of \source\ \citep{Kreykenbohm2004,LaBarbera2005,Suchy2012,Doroshenko2010,Furst2018}. Finally, to take into account the uncertainty in the absolute flux calibration between \Nu\ instruments, FPMA and FPMB, a multiplicative constant was added to the model which gives a normalization factor of 1.017. Consequently, the best-fit was obtained by a complex model constructed as {\scriptsize constant $\times$ tbabs $\times$ tbpcf $\times$ gabs $\times$ gabs $\times$ (bbodyrad + fdcut + gaussian)},  yielding $\chi^{2}_{\rm red}$ = 1.2 for 1808 degrees of freedom using 2019 \Nu\ observation. The broadband phase-averaged spectra of \source\ obtained by \Nu\ in 2019 and the corresponding residuals from the best-fit model are shown in Fig.~\ref{fig:spec}. The best-fit parameters and the corresponding uncertainties are also summarized in Table~\ref{tab:2}. 
 
There is no agreement on the value of the cutoff energy for \source\ as different models used to fit the spectra of the source yielded different values. Our best-fit model gives $E_{\rm cut}$ = 30.5$^{+0.3}_{-0.6}$ keV, which is consistent with the value $\sim29$ keV reported by \citet{Suchy2012} using the \textit{Suzaku} observation. \citet{Kreykenbohm2004} obtained $E_{\rm cut}$ = 10--15 keV using the Fermi-Dirac cutoff to fit the \textit{RXTE} data taken at the periastron passage. Other studies also reported the cutoff energy as $\sim$20 keV \citep[$RXTE$;][]{2004A&A...427..567M}, $\sim$20 keV \citep[$BeppoSAX$;][]{LaBarbera2005},  and $\sim$45 keV \citep[$NuSTAR$;][]{Furst2018}. The folding energy of $E_{\rm fold}$ = 5.04$^{+0.08}_{-0.19}$\,keV is also consistent with the results obtained by \citet{Kreykenbohm2004} and \citet{Furst2018}. 

We get the value for the local absorption $N_{\rm H,2}$ = 5.1 $\times~10^{23}$ cm$^{-2}$. It shows how deep the NS is engulfed by the companions wind. Our best-fit $N_{\rm H,2}$ value is higher than the one reported by \citet{LaBarbera2005} who used two {\sc wabs} models \citep{1983ApJ...270..119M} to account for  the photoelectric absorption. This inconsistency can be due to the different cross-section and continuum models used for fitting the spectra. We also confirm the presence
of the two absorption features reported in previous studies. We find that the first  line \citep[CRSF1; following definition from][]{Furst2018} is centered at 55.1 keV with a width of $\sigma_{\rm CRSF1}$ = 4.2 keV and the line depth of $D_{\rm CRSF1}$ = 8.5. The second line (CRSF2) is detected at 36.7 keV with $\sigma_{\rm CRSF2}$ and $D_{\rm CRSF2}$ of 11.8 keV and 42.1, respectively. The results are slightly different compared to previous studies \citep{Kreykenbohm2004,Furst2018} where the lines were reported at similar energies. We note, however, that findings reported in the literature are also controversial, e.g. a single feature at $\sim$46 keV was reported by \citet{LaBarbera2005} and \citet{Doroshenko2010}. We find that \Nu\ data are better described by a former model, and do not find any evidence for a $\sim46$\,keV feature.  

We applied our best-fit model also to the two archival \Nu\ observations taken in 2014 and 2015 in order to see the differences between the wind-fed and disk-fed states. The best-fit parameters for these two observations are listed in Table~\ref{tab:2} and the corresponding phase-averaged spectra are shown in Fig.~\ref{fig:spec}. The parameters for these two observations are broadly consistent with previous studies of the source. For the 2015 spectrum the fit detected the CRSF1 at 50.1 keV and CRSF2 at 35.8 keV;  both values are consistent with the results reported by \citet{Furst2018} who also used the Fermi-Dirac cutoff (\textsc{fdcut}) model. However, both are smaller than the values we reported for the disk-fed state spectra (\Nu\ 2019). Similarly, for the 2014 spectrum, the fit yielded the CRSF lines at 49.6 and 34.6 keV and the cutoff energy at 44.5 keV  consistent with the values reported by \citet{Furst2018}. However, our cutoff energy for 2015 observation is 31.0 keV which is smaller that what \citet{Furst2018} obtained for this observation. The $N_{\rm H,2}$ for 2015 and 2014 are $\sim$3.2 $\times~10^{23}$ and $\sim$4 $\times~10^{23}$ cm$^{-2}$, respectively, consistent with the results presented by \citet{Furst2018}, but both are smaller than the column density obtained for the 2019 spectra. The other parameters are more or less consistent with the values from the 2019 observation.

\begin{figure}
\begin{center}  
\includegraphics[width=\columnwidth]{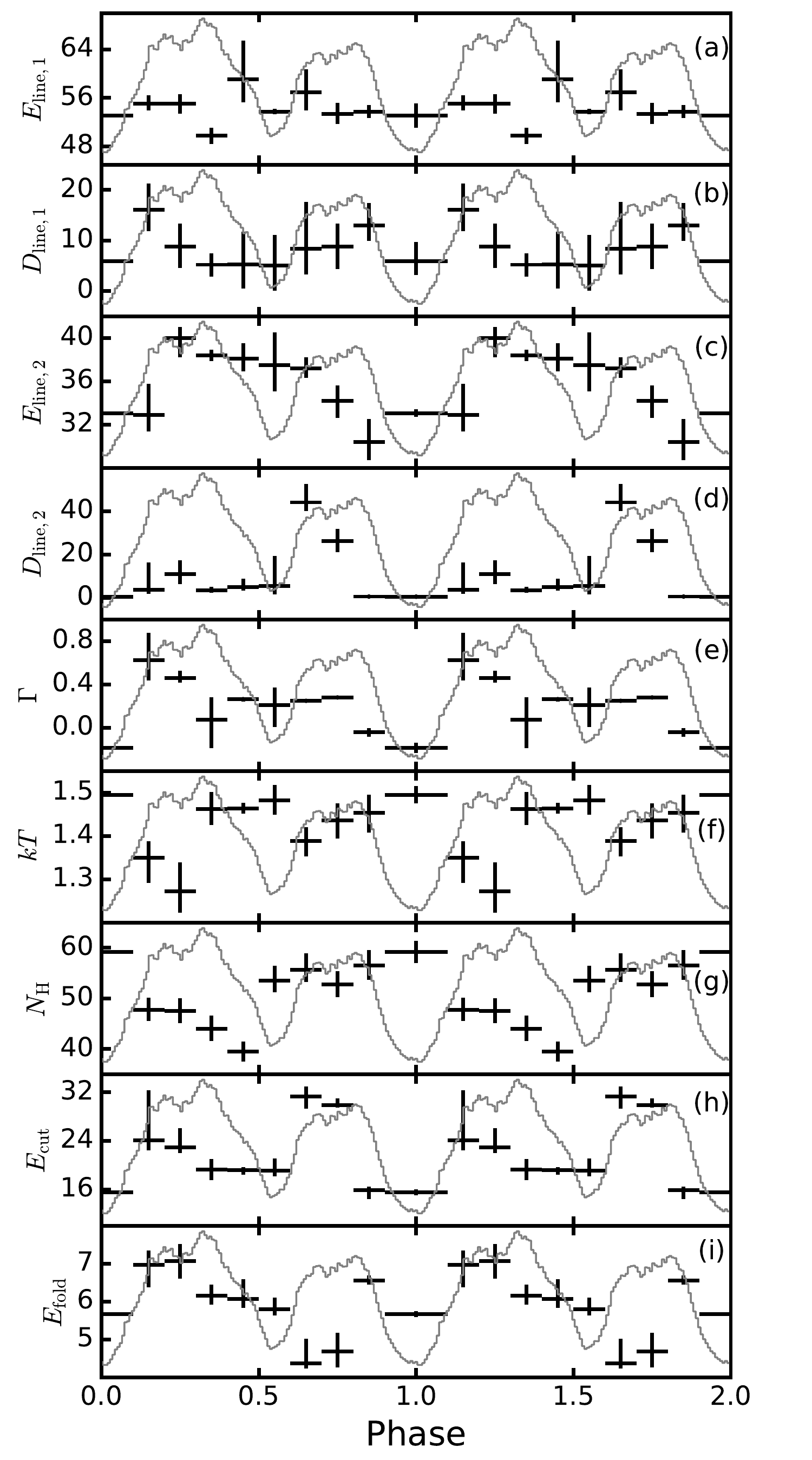}
\end{center}
\caption{Spectral model parameters variations as
a function of pulse phase. The crosses from upper panel to the lowest one show: (a) first cyclotron line energy (CRSF1), (b) first cyclotron line depth, (c) second cyclotron line energy (CRSF2), (d) second cyclotron line depth, (e) photon index, (f) blackbody temperature, (g) partial covering column density $N_{\rm H,2}$, (h) cutoff energy, and (i) folding energy. The grey line in each panel indicates the averaged pulse profile in the energy range 3--79 keV. } 
\label{fig:resolved} 

\end{figure}

\subsection{Phase-resolved spectroscopy}

Variations of the spectral parameters with pulse phase can provide useful insights about the physical conditions in the emission region. Therefore, we performed  phase-resolved spectroscopy for \source\ to study the evolution of the spectral parameters as a function of pulse phase. We split the spin period into 9 phase bins as indicated in Fig.~\ref{fig:resolved}. The width of the bins was chosen at $\Delta \phi$ = 0.1, except the ninth one which was chosen to be $\Delta \phi$ = 0.2 to ensure comparable counting statistics in all the bins. After extracting the spectra for each phase bin, we grouped them to have at least 1 count in each energy bin. We then fitted each spectrum with the same model we used to describe the averaged spectra and used the W-statistics \citep{Wachter1979} to describe the quality of the fit. Due to the limited statistics, we fixed the width of CRSF1 $\sigma_{\rm CRSF1}$ to the phase-averaged value as it was poorly constrained.

It is interesting to note that phase dependence of energies of the two absorption features are quite different. The position of the first CRSF remains almost constant at around 55 keV except for the main maximum of the pulse where it drops to $\sim$50 keV, i.e. it lies in the range 44--55 keV reported by \citet{Furst2018} for the 2015 \Nu\ observations. The depth of the line, $D_{\rm CRSF1}$, varies from 5 to 16 and the minimum is achieved in the fall of the main peak between the pulses. Energy of the second CRSF changes between 30 keV in the fall of the second pulse to the 40 keV in the middle of the main pulse. Similar behaviour for the low-energy CRSF was also reported by \citet{Kreykenbohm2004} and \citet{Suchy2012}. However, the trend is not consistent with \citet{Furst2018} where the CRSF is reported to be maximal at the rising part of the second pulse and minimal at the falling part of the same pulse. The line depth $D_{\rm CRSF2}$ varies mostly in the range of 0.5--12 except for the rising phase of the second pulse where it reaches to 44.

The photon index $\Gamma$ changes slightly from $-0.18$ in the main minimum to 0.62 near the main maximum. It decreases in the falling part the main pulse and remains almost constant during the second pulse and again decreases in the fall of the second pulse. These values roughly match the results by \citet{Kreykenbohm2004} ($\Gamma$ between $-0.2$ and 1.0) but are in contrast with the results of \citet{Suchy2012}, where $\Gamma$ varied in the range 0.5--1.3. The \Nu\ 2014 and 2015 data also showed photon index variations between 0.5--1.2 throughout the pulse \citep{Furst2018}. The cutoff energy is more variable, varying from $\sim$15~keV in the main minimum to $\sim$31~keV in the second pulse. The folding energy $E_{\rm fold}$ is slightly correlated with the pulse amplitude and shows a minimum of $\sim$4~keV in the rising part of the second peak and a maximum of 7~keV in the middle of the main peak. The temperature of the soft component is quite stable over the pulse with very small variations between 1.27 and 1.51 keV. The absorbing component $N_{\rm H,2}$ (from the clumped wind) varies in the range 4--6 $\times$ $10^{23}$\,cm$^{-2}$. The covering fraction as expected is quite consistent with the phase-averaged value and has very small variations around 0.84.

\section{Discussion} \label{sec:discussion}

\subsection{Spectral Features}

Our best-fit model for the latest \Nu\ observation during the 2019 spin-up episode implies that the overall spectral shape is similar to that previously reported in the literature for the wind-fed state. The detected CRSFs' energies of $\sim$55 and $\sim$37 keV are, however, slightly higher than the values obtained with the same model for observations outside of the spin-up episode ($\sim$50 and $\sim$36 keV respectively). In the energy range 3--79 keV, the absorbed X-ray fluxes for the three observations of the source in 2019, 2015 and 2014 observations are significantly different at 8.3$\times10^{-9}$, 2.5$\times10^{-9}$ and 1.6$\times10^{-9}$ erg s$^{-1}$cm$^{-2}$, respectively, so this difference in spectral shape might be simply associated with the source luminosity rather than the accretion state. That is, a higher CRSF energy corresponds to the higher flux, which might indicate a positive correlation between the luminosity and the CRSF energy. We note that \citet{LaBarbera2005} also found evidence for a positive CRSF energy--luminosity correlation for the source, which would imply a sub-critical regime of accretion in this source. \citet{Kreykenbohm2004} also suggested that accretion might be sub-critical, associating the observed dependence of the CRSF energy on pulse phase with rotation of the NS. 

On the other hand, \citet{Furst2018} mention that a positive luminosity-correlation of the CRSF can be explained if the deceleration happens in two stages, first in the radiation-dominated shock and closer to the NS surface  by Coulomb collisions. This, according to \citet{Becker2012}, would require a NS mass in excess of 1.8$M_{\odot}$. 
The  deviation in the ratio of energies of the two absorption features then have been attributed  to a difference in altitude by 1.4~km of the two emission regions, with the high-energy CRSF originating from the NS surface and the low-energy one from the accretion shock. In the 2019 observation likewise, the ratio of the energies of the two features does not appear to be harmonic, so the issue persists. \citet{Doroshenko2010} also invoked a tall accretion column to reconcile high magnetic field of the source deduced from the observed spin-evolution with the observed line energy which would be in odds with the sub-critical accretion. The presence of two non-harmonically related lines may be explained by  the off-set magnetic dipole \citep{Iyer2015}, which would naturally produce two poles of different magnetic field strengths.

\subsection{Disk or wind accretion?}

Mass transfer in the system has been suggested to occur both through homogeneous wind and a focused gas stream, which hits the NS close to the periastron and apastron, and shapes the observed orbital light curve \citep{Leahy1991}. \citet{Leahy2008} fitted the orbital light curve and found that the mass loss rate in the stream is a factor of two higher than in the companion wind, making the stream an important component. Their model required wind velocities from 400 to 600~km~s$^{-1}$ which are in agreement with spectral observations of \citet{Kaper2006}. Understanding how the mass transfer happens is important for determining the material torques experienced by the NS.  

The newly observed strong spin-up episode of \source\ is difficult to explain by direct wind accretion. Earlier, \citet{Koh1997} reported two strong spin-up events that took place in 1991 and 1993 and suggested that they were the result of a transient accretion disk forming around the NS and providing a steady acceleration torque. The spin-up episodes began shortly after the periastron passage and lasted from 10 to 30 days  which corresponds to a viscous timescale of the disk \citep{Koh1997}. The infrequent spin-up events cannot be primarily due to a tidal outflow at the periastron passage, but more likely related to episodes of enhanced mass-loss from the companion which would also lead to the associated increase in apastron flux \citep{Koh1997,Doroshenko2010}. The two later spin-up events in 2009 and 2010 seen in the GBM data were reported in \citet{2010ATel.2712....1F} and \citet{2011fxts.confE..43W}, where the connection to the apastron flare was noted as well. Considering the similarities, the 2019 event most likely has same origin, i.e. related to the formation of a transient accretion disk. This is supported by our examination of the disk formation in the light of \citet{Karino2019} model, which showed that the circularization radius of the wind matter can be larger than the magnetospheric radius in apastron, allowing for the disk to form. 
However, because the circularization radius is comparable to the magnetospheric radius, no extensive accretion disk can form, and accretion likely still mostly proceeds directly from wind. Indeed, the fact that the observed orbital light curve retained its typical features (i.e. pre-periastron and apastron flares) also suggests this scenario.  This may explain why the spectral and temporal properties of the wind and disk states are so similar. On the other hand, observed spin-up must be powered by the disk accretion, so we conclude that while most of the accreted mass is supplied by the wind, most of the angular momentum must be transferred via an accretion disc.

\begin{figure}
\begin{center}  
\includegraphics[width=\columnwidth]{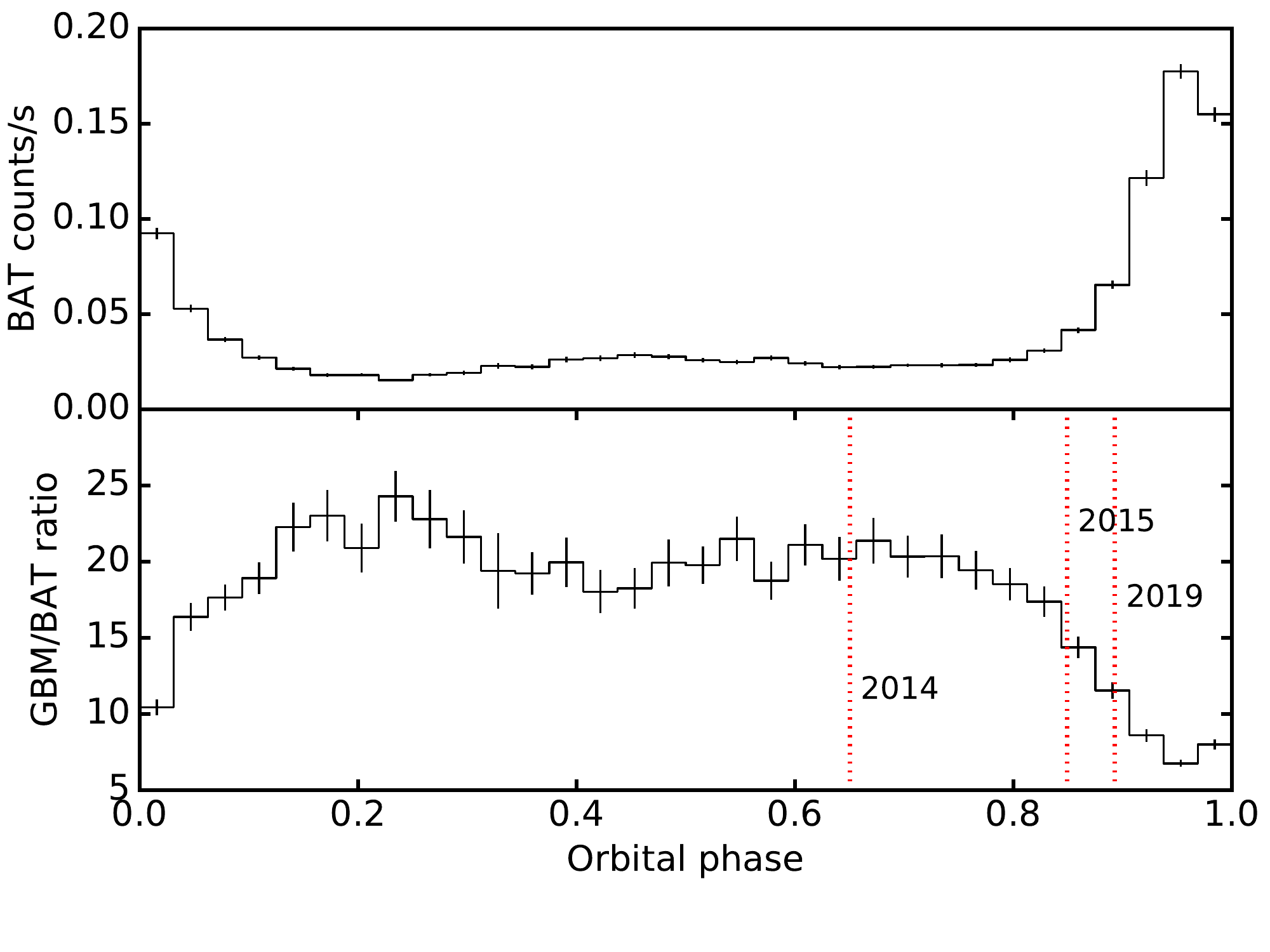}
\vspace{-1cm}
\end{center}
\caption{BAT flux (top panel) and GBM/BAT flux ratio (bottom panel) folded over the orbital period of the source (41.5 d). The red vertical dotted lines indicate the phases at which the \Nu\ observations were performed.}
\label{fig:ratio} 
\end{figure}

The difference in the pulse profiles shown in Fig.~\ref{fig:pp2} is a notable indicator of changes in the geometry of the emitting region with the binary orbital phase. The orbital variation of the pulse profile has been shown earlier by \citet{LaBarbera2005} and \citet{Evangelista2010}. The changes are also apparent in the energy dependence of the pulsed fraction in Fig.~\ref{fig:pf}. The dip at 6.3~keV is an iron feature which \citet{1991MNRAS.252..156T} explain as an indication of an extended emission region around the NS. Our comparison of the \Nu\ observations also shows that the pulsed fraction is lower closer to periastron, similarly to the result of \citet{2002ApJ...574..879E}. In principle, the presence of an accretion disk in 2019 could lower the pulsed fraction by increasing the emitting region on the NS surface due to deeper penetration to the magnetosphere. This is also supported by the size of emitting region on the NS surface in the disk-fed and wind-fed states obtained using the {\sc bbodyrad} normalization (see~Table~\ref{tab:2}). Assuming distance to the source $d$ = 3.53 kpc, the emission region radii were calculated to be  0.380$^{+0.001}_{-0.002}$, 1.00$^{+0.03}_{-0.05}$, and 2.25$^{+0.03}_{-0.02}$~km, for 2014, 2015 and 2019 observations, respectively.

However, during the near-periastron observation by \citet{2002ApJ...574..879E} there was no indication for the spin-up implying that the lower pulsed fraction must be unrelated to the disk. This can be also illustrated by comparison of the orbital light curves as observed by {\it Swift}/BAT and {\it Fermi}/GBM instruments. Indeed, GBM measures pulsed flux, whereas BAT the total flux in the hard energy band, therefore the ratio of these fluxes can be used as a measure of the pulsed fraction \citep{2018MNRAS.479L.134T}. The ratio of the folded orbital light curves from the two instruments is presented in Fig.~\ref{fig:ratio} along with the orbital phases at which the \Nu\ observations were performed (here the three spin-up episodes observed in the GBM era are excluded). Indeed, close to the periastron, the GBM/BAT ratio drops, which implies an overall drop of the pulsed fraction during the pre-periastron flare regardless on the presence of an accretion disk. This matches the observed pulsed fraction close to periastron. The lower pulsed fraction may be caused by scattering in the dense wind environment or obscuration by the extended gas stream, which is in agreement with the pulsed fraction remaining high at the highest energies of our band. Therefore, it is unclear how much, if at all, the disk affects the pulsed fraction.

\section{Conclusions}

In this work, we have presented a detailed investigation of the well-known wind-fed XRP \source\ using \Nu\ observation of the source during an unusual spin-up episode exhibited in January-March 2019. Because wind is not an effective source of torque to steadily accelerate the NS, it is believed that such spin-up episodes occur when a transient accretion disk forms around the NS. The start of the spin-up episode is between the periastron and apastron as with  previously reported spin-up events, which is in line with the formation of the accretion disk close to the apastron.  
Therefore, we used two archival \Nu\ observations which were obtained during the wind-fed state in 2014 and 2015 to investigate how the presence of an accretion disk affects the temporal and spectral properties. Comparison of the spectral properties of the source in two states revealed no major differences in the phase-averaged and the phase-resolved spectra between the disk- and wind-fed states beyond possible correlation of the CRSF energy with luminosity. The pulsed fraction was revealed to be lower during the disk-fed state, but that change is most likely related to the proximity to the periastron rather than a  change of the accretion mechanism. We also note  that evolution of the source flux with the orbital phase remained similar to that normally observed throughout the spin-up episode, although the apastron peak in the orbital light curve was significantly stronger than usual. Based on this fact, and the absence of notable changes in the observed spectral and timing properties of the source, we conclude that  bulk of the mass powering the observed X-ray emission was accreted directly from the wind also during the spin-up episode. On the other hand, accretion of significant angular momentum unambiguously points towards the disk accretion, so we finally conclude that accretion directly from the wind and through a transient accretion disk takes place during the spin-up episode. 

\begin{acknowledgements}
This work was supported by the Russian Science Foundation grant 19-12-00423 (SST, VD, SVM). 
This work was also supported by the grant from the Vilho, Yrj\"o and Kalle V\"ais\"al\"a Foundation of the Finnish Academy of Science and Letters (JM). 
We also acknowledge the support from the Academy of Finland travel grants 317552 (SST, JP),  324550 (SST) and 322779 (JP). The authors would like to acknowledge networking support by the COST Actions CA16214 and CA16104.
We grateful to the {\it NuSTAR} team for approving the DDT observation of \source.

\end{acknowledgements}

\bibliographystyle{aa}
\bibliography{library}

\end{document}